\documentclass[pra,nofootinbib,twocolumn, superscriptaddress]{revtex4-2}
\usepackage{lipsum}
\usepackage{amsmath,amssymb,bm,graphicx,comment,mathrsfs}
\usepackage{multirow}

\usepackage{dcolumn}
\usepackage{latexsym}
\usepackage[colorlinks=true, linkcolor=blue]{hyperref}
\usepackage[normalem]{ulem}

\newcommand{\Tt}{\Theta}

\newcommand{\ket}[1]{| #1 \rangle}
\newcommand{\bracket}[2]{\langle #1 | #2 \rangle}
\newcommand{\ketbra}[2]{| #1 \rangle \langle #2 | }
\newcommand{\tr}{\mathop{\mathrm{tr}}\nolimits} 	
\newcommand{\HA}{\mathop{\mathcal{H}}\nolimits} 
\newcommand{\KA}{\mathop{\mathcal{K}}\nolimits} 
\newcommand{\LA}{\mathop{\mathcal{L}}\nolimits} 
\newcommand{\SA}{\mathop{\mathcal{S}}\nolimits} 
\newcommand{\IA}{\mathop{\mathcal{I}}\nolimits}
\newcommand{\R}{\mathop{\mathbb{R}}\nolimits} 
\newcommand{\I}{\mathop{\mathbb{I}}\nolimits}
\newcommand{\E}{\mathop{\cal{E}}\nolimits} 
\newcommand{\Ketbra}[2]{| #1 \rangle\!\rangle \langle\!\langle #2 |}
\newcommand{\Bracket}[2]{\langle\!\langle  #1 | #2 \rangle\!\rangle}

\newtheorem{theorem}{Theorem}
\newtheorem{definition}[theorem]{Definition}
\newtheorem{lemma}[theorem]{Lemma}

\newtheorem{proposition}[theorem]{Proposition}

\begin{document}

\sloppy

\title{Construction of general symmetric-informationally-complete–positive-operator-valued measures by using a complete orthogonal basis}

\author{Masakazu Yoshida}
\email{yoshida@ise.osaka-sandai.ac.jp}
\affiliation{
Faculty of Design Technology, Osaka Sangyo University, 
\\3-1-1 Nakagaito, Daito-shi, Osaka, 574-8530, Japan
}

\author{Gen Kimura}
\email{gen@shibaura-it.ac.jp}
\affiliation{
College of Systems Engineering and Science, Shibaura Institute of Technology, 
\\307 Fukasaku, Minuma-ku, Saitama-shi, Saitama, 337-8570, Japan.
}

\date{\today}

\begin{abstract}
A general symmetric informationally complete (GSIC)-positive operator valued measure (POVM) is known to provide an optimal quantum state tomography among minimal IC-POVMs with a fixed average purity. 
In this paper, we provide a general construction of a GSIC-POVM by means of a complete orthogonal basis (COB), also interpreted as a normal quasiprobability representation. 
A spectral property of a COB is shown to play a key role in the construction of SIC-POVMs and also for the bound of the mean-square error of the state tomography. 
In particular, a necessary and sufficient condition to construct a SIC POVM for any $d$ is constructively given by the power of traces of a COB. 
We give three simple constructions of COBs from which one can systematically obtain GSIC POVMs.  
\end{abstract}


\maketitle

\section{Introduction}\label{intro}
An appropriate quantum state preparation rapidly increases in its importance 
according to the development of applications in quantum information theory 
such as quantum computation and quantum key distribution. 
An intended quantum effect can be obtained when quantum states used in such applications are not disturbed. 
Therefore, it is important to experimentally check whether the quantum system is appropriately prepared. 
Quantum state tomography provides a way to determine completely quantum states with their statistical information. 

An informationally-complete (IC)-positive-operator-valued measure (POVM) \cite{P77,BL89,Bu91,DPS04} is suitable for linear quantum state tomography 
since any quantum state can be determined completely by its measurement statistics. 
Any IC POVM for a $d$-level quantum system has at least $d^2$ POVM elements, whence an IC POVM with $d^2$ POVM elements is called minimal. 
A quantum measurement represented by a symmetric-informationally-complete (SIC) POVM \cite{RBSC04}
is known to be optimal for linear quantum state tomography \cite{Sc06,Zh14_2}. 
However, the existence of SIC POVMs has been shown in limited dimensions \cite{SG10,ACFW17,Sc17} and 
it remains an open question whether the SIC exists in all dimensions. 
For the most up to date information, see, for example, \cite{FHS17}.

A general SIC POVM \cite{Ap07,GK14} is a generalization of a SIC POVM. 
Different from a SIC case, POVM elements in a GSIC POVM are not necessarily of rank $1$ and the existence of GSIC POVMs has been shown in all dimensions \cite{Ap07,GK14}. 
Zhu has shown \cite{Zh14} that a GSIC POVM provides an optimal measurement for the linear quantum state tomography among minimal IC POVMs with a given average purity of a POVM. 
Uncertainty relations of GSIC POVMs are studied in different contexts such as the entropic uncertainty relation \cite{Ra14}, the uncertainty and complementarity relation using generalized Wigner-Yanase-Dyson skew information \cite{HWF20}, 
and the improved state-dependent entropic uncertainty relation \cite{HCW21}. 
Entanglement detection using the index of coincidence for GSIC POVMs as well as its experimental implementation has also been studied in \cite{CLF15,XZZ16,SLLF18,LLFW18,LC21}. 

In this paper we characterize GSIC POVMs by using a complete orthogonal basis (COB) of the set of Hermitian operators. 
The conditions of informational completeness, symmetry, and completeness (normalization) of POVMs are derived directly from the properties of COBs. 
We observe that a spectrum property of COBs plays a key role in the construction of a SIC POVM and also determines the bound of the scaled mean-square errors of the minimal IC POVMs with a given average purity. 
In particular, any canonically constructed GSIC POVM is shown to give a SIC POVM for a qubit system, while for higher-level systems, conditions that yield SIC POVMs are given by the conditions for the power of traces of a COB. 
We also provide three simple constructions of COBs (and hence those of GSIC POVMs) from any sub orthonormal operator basis and also from a complete set of mutually unbiased bases \cite{Iv81,WF88}. 
Incidentally, the notion of a COB can be interpreted as the normal quasiprobability representation (NQPR) studied in \cite{Zh16}. 
Hence, our constructions of COBs also serve as those of NQPRs. 

This paper is organized as follows. In Sec. \ref{pre} we review GSIC POVMs in a slightly wider context. 
In Sec. \ref{Ai}, we introduce a COB and investigate its spectral properties. 
In particular, we give a construction of GSIC POVMs by means of COBs. 
In Sec. \ref{constnb} we give three constructions of COBs. 
We summarize this paper in Sec. \ref{sum}.

\section{Preliminaries}\label{pre}
Throughout the paper, $\HA$ is a finite dimensional Hilbert space with dimension $d \ge 2$ and $\LA(\HA)$ is the $d^2$-dimensional Hilbert space of linear operators on $\HA$ with respect to the Hilbert-Schmidt inner product. 
For both Hilbert spaces, we use the Dirac notation with single or double angular brackets as follows: Inner products on $\HA$ and $\LA(\HA)$ are denoted by the angular brackets $\bracket{\psi}{\phi} \ (\psi,\phi \in \HA)$ and the double angular brackets $\Bracket{A}{B} = \tr A^\dagger B \ (A,B \in \LA(\HA))$, respectively. 
The operator $\ketbra{\psi}{\phi}$ and the super operator $\Ketbra{A}{B}$ are also used in a conventional sense, e.g., $\Ketbra{A}{B} C := \Bracket{B}{C} A$. 
The set of density operators, i.e., positive operators with unit trace, is denoted by $\SA (\HA):=\{\rho\in\LA (\HA) \mid \rho \ge 0, \tr\rho =1 \}$. 

Let $F = (F_i)_{i=1}^{n}$ be a discrete POVM on $\HA$, i.e., $F_i\geq 0$ for any $i$ and $\sum_{i=1}^{n} F_i=\I$ where $\I$ is the identity operator.  
Here $F$ is called an informationally complete POVM if the statistics of the measurement of $F$ determine the underlying quantum state. 
In other words, $F$ is an IC-POVM if for $\rho, \sigma\in\SA (\HA)$, $\tr F_i\rho = \tr F_i\sigma \ (\forall i = 1,2,\ldots,n)$ implies $\rho=\sigma$. 
One can show that a POVM is IC if and only if it spans $\LA(\HA)$. (For the readers' convenience, we give a simple proof for this fact in Appendix \ref{spICPOVM}.) 
An IC POVM is thus called minimal if $n=d^2$. 

A rank $1$ POVM $F= (F_i=\ketbra{\psi_i}{\psi_i})_{i=1}^{d^2}$ is called a symmetric-informationally-complete POVM \cite{RBSC04} if it satisfies 
\begin{eqnarray*}
\tr F_i^2 &=& ||\psi_i||^4 = a \quad \forall i, \\
\tr F_i F_j  &=& |\bracket{\psi_i}{\psi_j}|^2 = b \quad \forall i \neq j, 
\end{eqnarray*}
where $a$ and $b$ are constants dependent only on the dimension $d$. 
Note that these constants are automatically determined as $a = \frac{1}{d^2}$ and $b = \frac{1}{d^2(d+1)}$. (This is shown by taking traces over the equations $\I = \sum_i F_i = \sum_i \ketbra{\psi_i}{\psi_i}$ and $\I = (\sum_i F_i)^2$.) 
Moreover, one can show that a SIC POVM spans $\LA(\HA)$ (see the general argument below) and hence is informationally complete and minimal.   
However, the existence of SIC POVMs for an arbitrary dimension is a long standing open problem and has only been shown analytically (or numerically) in limited dimensions (see, e.g., \cite{FHS17}). 

A natural generalization of a SIC POVM is given by relaxing the condition for the rank: A POVM $(G_i)_{i=1}^{d^2}$ is called a general SIC POVM \cite{Ap07,GK14} if it satisfies 
\begin{eqnarray*}
\tr G_i^2 &=& a' \quad \forall i, \\
\tr G_iG_j &=& b' \quad \forall i\neq j, 
\end{eqnarray*} 
where $a'$ and $b'$ are constants dependent only on $d$. 
Different from a SIC POVM, the existence and the construction of GSIC POVMs have been shown in all dimensions \cite{Ap07,GK14}. 

Here we review some of the properties of a GSIC POVM by further generalizing the number of POVM elements to be arbitrary $n$: $G=(G_i)_{i=1}^{n}$. First, the parameters $a'$ and $b'$ are not independent and satisfy
\begin{equation}\label{eq:rela'b'}
a' + (n-1)b' = \frac{d}{n}.  
\end{equation}
This is seen by observing $d=\tr \I^2=\tr \{(\sum_iG_i)(\sum_jG_j) \}=\tr \{\sum_i(G_i^2+\sum_{i\neq j}G_iG_j)\}=n a'+n(n-1)b'$. This also determines the trace of $G_i$:
\begin{equation}\label{eq:trG}
\tr G_i = \tr G_i \bigg(\sum_j G_j\bigg) = a' + (n-1) b' = \frac{d}{n}. 
\end{equation}
Secondly, the parameter $a'$ satisfies 
\begin{equation}\label{eq:aineq}
\frac{d}{n^2} < a' \le \frac{d^2}{n^2}.  
\end{equation}
The first inequality follows from the Schwarz inequality: $\frac{d}{n} = \tr G_i = \tr G_i \I \le \sqrt{\tr G_i^2} \sqrt{\tr \I^2} = \sqrt{a'}\sqrt{d}$. However, the equality implies that $G_i = \frac{d}{n}\I$ and is excluded in order to keep the IC condition. 
The second inequality is shown by the following elementary fact: For any positive operator $A \ge 0$, 
\begin{equation*}
\tr A^2 \le (\tr A)^2,
\end{equation*}
where the equality holds if and only if $A$ is a rank $1$-operator. Applying each $G_i \ge 0$ shows the second inequality and also, in the case $n=d^2$, the equality holds if and only if $G$ is a SIC. 
Finally, it holds that $G$ is IC if and only if $n \ge d^2$. To see this, it is enough to show that $G$ is linearly independent (hence $n \ge d^2$ if and only if $G$ spans $\LA(\HA)$). Suppose that $\sum_{i=1}^{n} x_iG_i=0$ for complex numbers $x_i$. 
By taking the trace over the equation and using Eq. \eqref{eq:trG}, one has $\sum_i x_i=0$. 
Next, multiplying $G_j$ to $\sum_{i=1}^{n} x_iG_i=0$ and taking its trace shows $0=\sum_i x_i\tr G_iG_j =  (a'-b') x_j + b' \sum_i x_i$. 
Combining these results, one gets $x_j=\frac{b'}{b'-a'}\sum_i x_i = 0$ for all $j$. 
Note here that $b' \neq a'$ otherwise \eqref{eq:rela'b'} implies $a' = \frac{d}{n^2}$ violating the first inequality in \eqref{eq:aineq}.  
In the following, we consider the most interesting case $n = d^2$. 

In the linear quantum state tomography, Zhu \cite{Zh14} revealed the tomographic significance of GSIC POVMs in the following sense. For any IC POVM measurement $(\Pi_i)_{i=1}^n$, there exists a set of operators $(\Tt_i)_{i=1}^n$ with which any density operator $\rho$ can be written as $\rho = \sum_{i=1}^np_i\Tt_i$ where $p_i:=\tr\Pi_i\rho$ is the probability to get the $i$th outcome of the POVM $(\Pi_i)_{i=1}^n$ under the state $\rho$. 
Let $f^{(N)}_i$ be the frequency to get the $i$th outcome by the individual measurement of $(\Pi_i)_{i=1}^n$ under $N$ copies of $\rho$. Then, a natural estimated state is given by $\hat \rho^{(N)}=\sum_{i=1}^nf^{(N)}_i\Tt_i$. 
The scaled mean squared error (MSE) $\E(\rho)$ is defined by the expectation value of the error $\frac{\|\rho - \hat{\rho}^{(N)} \|}{N}$, where $\|\cdot\|^2 := \Bracket{\cdot}{\cdot}$ is the Hilbert-Schmidt norm. 
One can show that $\E(\rho)=\sum_{i=1}^np_i\tr \Tt_i^2 -\tr\rho^2$ \cite{Sc06}.  

In \cite{Zh14}, Zhu had shown that, among minimum IC POVMs with the fixed average purity (see below), the maximal scaled MSE $\E_{\max}(\rho)$ over all pure states (more generally over all unitary equivalent states) is bounded from below as  
\begin{equation}\label{mMSE_GSIC}
\E_{\max}(\rho)\geq \frac{(d^2-1)^2}{d^2\wp -d} + \frac{1}{d} -\tr\rho^2.
\end{equation}
Here, $\wp$ is the average purity of an IC-POVM $(\Pi_i)_{i=1}^{n}$ defined by 
\begin{equation}\label{eq:apu}
\wp: = \sum_i\wp_i \frac{\tr \Pi_i}{d}
\end{equation}
where $\wp_i:=\frac{\tr\Pi_i^2}{(\tr\Pi_i)^2}$ is the purity of $\Pi_i$. 
Interestingly, Zhu had shown that the minimum of \eqref{mMSE_GSIC} is attained if and only if the IC-POVM is a GSIC POVM. 
Therefore, one can consider a GSIC POVM as an optimal measurement among all minimal IC-POVMs with fixed average purity that minimize the scaled MSE for the worst case scenario of states. 

\section{Characterization of GSIC by complete operator basis}\label{Ai}
In the following, we consider the real Hilbert space of the set of all Hermitian operators $\KA = \{ A \in \LA(\HA) \ | \ A = A^\dagger\}$. 
Let us start by introducing a useful operator basis for $\KA$.  
\begin{definition}
An operator basis $(A_i)_{i=1}^{d^2}$ for $\KA$ is called a complete orthogonal basis if it satisfies
\begin{itemize}
\item[(i)] Sub-orthonormality: $\Bracket{A_i}{A_j} = \frac{1}{d} \delta_{ij}$,
\item[(ii)] Completeness: $\sum_i A_i =\I$. 
\end{itemize}
\end{definition}
See Appendix \ref{app:obwfs} for examples. 
Note that the normalization constant $\frac{1}{d}$ in (i) is automatically determined by (ii).  
By completeness, one has $\tr A_i = \tr A_i (\sum_j A_j) = \frac{1}{d}$. 
If there is a positive element $A_i \ge 0$ for some $i$, $\frac{1}{d}= \tr A_i^2 \le (\tr A_i)^2 = \frac{1}{d^2}$, contradicting $d \ge 2$. 
Therefore, any element $A_i$ of a COB cannot be positive, and the minimum eigenvalue of each $A_i$ is strictly negative. 
Here, we define an important value for a COB: 
\begin{equation*}\label{eq:lamast}
\lambda^\ast := \frac{1}{1+d^2 \tau},
\end{equation*}
where 
\begin{equation*}\label{tau}
\tau : = \max_{i=1,2,\ldots,d^2} \{ |m_i| \mid  m_i : \text{the minimum eigenvalue of} \ A_i \}
\end{equation*}

\begin{proposition}\label{lam_bound}
The value $\lambda^\ast$ satisfies
\begin{equation}
\lambda^\ast \le \frac{1}{\sqrt{d+1}}. \label{lam}
\end{equation}
The upper bound is saturated if and only if all $A_i$ have the same eigenvalues: $\frac{(d-1)\sqrt{d+1}+1}{d^2} > 0$ with multiplicity $1$ and $\frac{1-\sqrt{d+1}}{d^2} < 0$ with multiplicity $d-1$. 
\end{proposition}
This is shown by the following lemma. 
\begin{lemma}\label{lem:xd}
Let $x_i \ (i=1,2,\dots,d)$ be $d \ge 2$-tuples of a real number in descending order with constraints 
\begin{equation*}
(i) \ \sum_i x_i = \frac{1}{d}, \quad (ii) \ \sum_i x_i^2 =  \frac{1}{d}. 
\end{equation*}
Then, the minimum $x_d < 0$ and satisfies 
\begin{equation}
|x_d| \ge \frac{\sqrt{d+1}-1}{d^2}. \label{eq:ineq}
\end{equation}
The bound is saturated if and only if 
\begin{eqnarray}
x_1 &=&\frac{1+ (d-1)\sqrt{d+1}}{d^2}, \label{x1}\\
x_2,x_3,\ldots ,x_{d} &=& \frac{1-\sqrt{d+1}}{d^2}. \label{x2_d}  
\end{eqnarray}
\end{lemma}
See Appendix \ref{pr:xd} for the proof. 
\\
{\it Proof of Proposition \ref{lam_bound}}. 
Note that the inequality \eqref{lam} is equivalent to 
\begin{equation*}\label{eq:ineqtau} 
\tau \ge \frac{\sqrt{d+1}-1}{d^2}.
\end{equation*}
However, this is shown to hold by applying Lemma \ref{lem:xd} to each eigenvalue of $A_i$ (noting that $\tr A_i = \frac{1}{d}$ and $\tr A_i^2 = \frac{1}{d}$). 
The equality condition also follows directly from one for \eqref{eq:ineq} in Lemma \ref{lem:xd}. 
\hfill $\blacksquare$
 
Now we provide a construction of a GSIC POVM by showing the connection between a GSIC POVM and a COB. 
\begin{theorem}\label{const:genSIC}
For any COB $(A_i)_{i=1}^{d^2}$ and $\lambda \in (0, \lambda^\ast] $, 
\begin{equation}\label{DefGk}
G_i = \lambda A_i + (1-\lambda) \frac{\I}{d^2} 
\end{equation}
forms a GSIC POVM. Conversely, for any GSIC POVM $(G_i)_{i=1}^{d^2}$ with constants $a'$ and $b'$, $(A_i)_{i=1}^{d^2}$ given by \eqref{DefGk} with $\lambda = \sqrt{1 - b'd^3}=\sqrt{\frac{d^3a'-1}{d^2-1}}$ forms a COB.  
\end{theorem}
{\it Proof.} Letting $(A_i)_{i=1}^{d^2}$ be a COB and $\lambda \in (0, \lambda^\ast] $, we show that 
$(G_i)_{i=1}^{d^2}$ of the form \eqref{DefGk} is a GSIC POVM.  
The completeness $\sum_iG_i = \I$ follows from that of $(A_i)_i$.  
Next $G_i $ is positive if and only if $\lambda m_i + (1-\lambda) \frac{1}{d^2} \ge 0$ where $m_i$ is the minimum eigenvalue of $A_i$. 
Since $m_i < 0$ as mentioned above, the condition is equivalent to $\frac{1}{1 + d^2 |m_i|} \ge \lambda$. 
This holds since $\lambda \in (0, \lambda^\ast] $, so we have $G_i \ge 0$. 
Moreover, the symmetric property of $(G_i)_i$ follows as    
\begin{eqnarray*}
\tr G_i G_j &=& \tr \Big\{\lambda A_i + (1-\lambda) \frac{\I}{d^2}\Big\}\Big\{\lambda A_j + (1-\lambda) \frac{\I}{d^2}\Big\} \\ 
&=& \lambda^2 \tr A_iA_j + \frac{(1-\lambda)\lambda}{d^2} \tr A_i \\
&& +\frac{\lambda(1-\lambda)}{d^2} \tr A_j + \frac{(1-\lambda)^2}{d^4} \tr \I \\
&=& \lambda^2 \frac{\delta_{ij}}{d} + \frac{1-\lambda^2}{d^3}. 
 \end{eqnarray*}	
Hence, $(G_i)_i$ is a GSIC POVM with the constants  
\begin{equation}\label{eq:ab}
a' = \frac{\lambda^2}{d} + \frac{1-\lambda^2}{d^3}, \ b' = \frac{1-\lambda^2}{d^3}. 
\end{equation}   
Conversely, letting $(G_i)_i$ be a GSIC with constants $a'$ and $b'$, we show that $A_i := \frac{1}{\lambda} (G_i - \frac{1-\lambda}{d^2} \I )$ forms a COB with $\lambda = \sqrt{1 - b'd^3}=\sqrt{\frac{d^3a'-1}{d^2-1}}$ (recalling the relation \eqref{eq:rela'b'} where $n =d^2$.) 
Using the symmetry $\tr G_i G_j = a' \delta_{ij} + (1-\delta_{ij})b'$ and $\tr G_i = \frac{1}{d}$, we have 
\begin{eqnarray*}
\tr A_i A_j &=& \frac{1}{\lambda^2}\tr \Big(G_i - \frac{1-\lambda}{d^2} \I\Big)\Big(G_j - \frac{1-\lambda}{d^2} \I\Big) \\
&=& \frac{1}{\lambda^2} \Big\{ (a'-b') \delta_{ij} + b' - \frac{2 (1-\lambda)}{d^3} + \frac{(1-\lambda)^2}{d^3}\Big\} \\
&=& \frac{1}{\lambda^2} \Big\{ (a'-b') \delta_{ij} + b' -\frac{1-\lambda^2}{d^3}\Big\} = \frac{1}{d} \delta_{ij}.  
\end{eqnarray*}
Finally, the completeness of $(A_i)_i$ follows from that of $(G_i)_i$. 
\hfill $\blacksquare$

Theorem \ref{const:genSIC} shows that any GSIC POVM including a SIC POVM can be constructed by a COB which is rather easy to construct (see the next section). 
Note that another construction of a GSIC POVM was given in \citep{GK14}. 
However, their construction needs two asymmetrical expressions, thereby it unnecessarily breaks a symmetry of a GSIC POVM in appearance. 
In contrast, our construction \eqref{DefGk} consists of a single expression; hence it does not introduce any redundant asymmetry. 

Before giving its construction, let us discuss the relation between a SIC POVM and a COB. 
Although there is freedom for the choice of $\lambda \in (0,\lambda^\ast]$, the extreme choice $\lambda = \lambda^\ast$ plays a crucial role in constructing SIC POVMs. 
In the following, we call such construction a canonical construction. 
Note that, by \eqref{eq:ab}, $(G_i)_i$ is a SIC POVM, i.e., $a' = \frac{1}{d^2}$, if and  only if $\lambda = \frac{1}{\sqrt{d+1}}$. 
Therefore, Proposition \ref{lam_bound} leads to the following proposition. 
\begin{proposition}\label{Lam_SIC}
A GSIC POVM canonically constructed by a COB (i.e., $\lambda = \lambda^\ast$) is a SIC POVM 
if and only if any one of the following conditions is satisfied: 
\begin{itemize}
\item[(i)] The upper bound of $\lambda^*$ in \eqref{lam} is saturated. 
\item[(ii)] $\tau = \frac{\sqrt{d+1}-1}{d^2}$ holds. 
\item[(iii)] All $A_i$ have the same eigenvalues: $\frac{1 + (d-1)\sqrt{d+1}}{d^2} > 0$ with multiplicity $1$ and 
$\frac{1-\sqrt{d+1}}{d^2} < 0$ with multiplicity $d-1$.  
\end{itemize}
\end{proposition}

The following result shows that any canonical construction in $d=2$ gives a SIC POVM:
\begin{proposition}\label{prop:t1}
For $d=2$, a canonical construction always gives a SIC POVM. 
\end{proposition}
{\it Proof.} Let $(A_i)_{i=1}^4$ be a COB. 
The eigenvalue equation for each $A_i$ reads $0 = \det(m \I - A_i) = m^2 - (\tr A_i) m +\frac 1 2 \{ (\tr A_i)^2-\tr A_i^2\}$. 
Therefore, $\tr A_i=\tr A_i^2=\frac{1}{2}$ implies that all eigenvalues of $A_i$ are the same $m=\frac{1\pm\sqrt 3}{4}$ satisfying condition (iii) in Proposition \ref{Lam_SIC}. 
\hfill $\blacksquare$ 

Note, however, that not all canonical constructions in the case $d \ge 3$ give SIC POVMs since higher contributions of $\tr A_i^n \ (3 \le n \le d)$ appear in the eigenvalue equations. However, we have the following proposition. 
\begin{proposition}\label{prop:t2} 
For any $d\ge 3$, the necessary and sufficient conditions for a canonical construction to give a SIC POVM are systematically derived: 
To be specific, the conditions are $\tr A_i^3 = \frac{31}{243}$ for $d=3$, $\tr A_i^3 = \frac{1}{512} \left(23+15 \sqrt{5}\right)$ and $\tr A_i^4 = \frac{1}{2048}(77 + 15 \sqrt{5})$ for $d=4$, etc. 
\end{proposition}
{\it Proof} By using Newton's identity (see e.g., \cite{ref:NI}), one can derive the characteristic equations for $A_i$ bearing in mind the constraints $\tr A_i=\tr A_i^2=\frac{1}{3}$. 
For example, for $d=3$, 
\begin{eqnarray*}
0 = \det(m \I - A_i)  = m^3 - \frac{1}{3} m^2 - \frac{1}{9} m - \frac{54 \tr A_i^3 - 8}{162}.  
\end{eqnarray*}
Therefore, by condition (iii) in Proposition \ref{Lam_SIC} for $d=3$, the necessary and sufficient condition for a canonical construction to give a SIC POVM is that all $A_i$ satisfy $\tr A_i^3 = \frac{31}{243}$. One can obtain the conditions similarly for any $d$.  
$\blacksquare$

The following proposition gives the physical meaning of the parameter $\lambda$ of a canonically constructed GSIC POVM in the context of quantum state tomography. 
\begin{proposition}\label{Coro_MSE_GSIC}
The average purity of a GSIC POVM constructed by a COB is given by $\wp=\frac{1}{d}\{ (d^2-1)\lambda^2+1\}$. 
The maximal scaled MSE for the GSIC POVM satisfies 
\begin{eqnarray*}
\E_{\max}(\rho) &=& \frac{d^2-1}{d}\frac{1}{\lambda^2}+\frac{1}{d}-\tr\rho^2 \\ 
&\geq& \frac{d^2-1}{d}(1+d)+\frac{1}{d}-\tr\rho^2. 
\end{eqnarray*}
The inequality is saturated if and only if the upper bound of $\lambda^*$ in \eqref{lam} is saturated which implies that the GSIC POVM is a SIC POVM.  
\end{proposition}
{\it Proof.} 
A direct computation of \eqref{eq:apu} for a GSIC shows $\wp=d^2a'$ hence by \eqref{eq:ab}, one obtains $\wp = \frac{1}{d}\{ (d^2-1)\lambda^2+1\}$. 
The first equality is the direct application of Zhu's result \eqref{mMSE_GSIC}. 
The second inequality follows from \eqref{lam} and $\lambda \in (0,\lambda^\ast]$. 
Finally, the last statement is shown by Proposition \ref{Lam_SIC}. 
\hfill $\blacksquare$

Hence, the larger the parameter $\lambda$, the less the maximal scaled MSE $\E_{\max}(\rho)$ 
and a SIC $(\lambda = \lambda^\ast = \frac{1}{\sqrt{d+1}}$) gives the minimal $\E_{\max}(\rho)$. 

Finally, we remark that a COB $(A_i)_i$ was used by Zhu \cite{Zh16} as an NQPR where a quantum state $\rho$ is represented by a (possibly negative) quasiprobability $\mu_i(\rho) = \tr A_i \rho$. 
The negativity of a COB $(A_i)_i$ is naturally defined by 
\begin{equation*}
N(\{A_i\}) := \max_{\rho \in \SA(\HA)} N(\rho)
\end{equation*}
where $N(\rho) := d \max \{0, - \min_i \mu_i(\rho)\}$.  
Theorem 1 in \cite{Zh16} shows a bound of the negativity where the bound is saturated if and only if  a POVM corresponding to the NQPR is a SIC POVM. 
In this context, Zhu also observed essentially the same results as Proposition \ref{lam_bound} and Proposition \ref{Lam_SIC} because one can readily show that 
\begin{equation}\label{Tau_Neg}
\tau = \frac{1}{d}N(\{A_i\}). 
\end{equation}
See Appendix \ref{pr:TauNeg} for the proof. 
Note that combination of the relation \eqref{Tau_Neg} and Proposition \ref{Coro_MSE_GSIC} for the canonically constructed GSIC POVM yields the following relation between the maximal scaled MSE for the GSIC POVM and the negativity: 
\begin{eqnarray*}
\E_{\max}(\rho) &=& \frac{d^2-1}{d}\{1+dN(\{A_i\})\}^2+\frac{1}{d}-\tr\rho^2 \\ 
&\geq& \frac{d^2-1}{d}(1+d)+\frac{1}{d}-\tr\rho^2. 
\end{eqnarray*}

In the next section we give several constructions of COBs for the construction of GSIC POVMs, 
which also serve as a construction of NQPRs in Zhu's context.  

\section{Constructions of a complete orthogonal basis}\label{constnb}

In this section, we provide several constructions of COBs. 
The general ideas of Constructions 1 and 2 are explained in Appendix \ref{app:obwfs} in more general settings. 
Construction 3 is based on the ideas developed in \cite{Wo06} and \cite{HHH205}. 

\bigskip 

\noindent {\bf Construction 1.} {\it 
With any orthonormal basis $(T_i)_{i=0}^{d^2-1}$ for $\LA(\HA)$ where $T_0 = \frac{\I}{\sqrt{d}}$ 
[i.e., a generator of su$(d)$], and any orthogonal $d^2\times d^2 $ real matrix $O = [O_{ij}]_{i,j=0}^{d^2-1}$ satisfying 
\begin{equation*}\label{eq:OProp}
O_{0 j} = \frac 1 d \quad (j = 0,1,\ldots,d^2-1), 
\end{equation*}
\begin{equation*}\label{eq:const1op}
A_i := \frac{1}{\sqrt{d}} \sum_{j=0}^{d^2-1} O_{ji} T_j \quad (i=0,1,\ldots,d^2-1)
\end{equation*}
is a COB. }

Note that both $(T_i)$ and $O$ are easily prepared, e.g., by using Gram-Schmidt orthogonalization starting from $\I$ and $(1,1,\ldots,1)^T \in \R^{d^2}$, respectively.  
Importantly, any COB can be obtained through this construction. 
See Appendix \ref{app:obwfs} for details in more general settings. 

\bigskip 

The next construction only uses a generator of su$(d)$; hence it is more economic and concrete than the first construction at the cost of losing generality. 

\noindent {\bf Construction 2.} {\it 
Let $(T_i)_{i=0}^{d^2-1}$ be an orthonormal basis for $\LA(\HA)$ with $T_0 = \frac{\I}{\sqrt{d}}$. 
Construct an orthonormal basis $(S_i)_{i=0}^{d^2-1}$ by the Gram-Schmidt orthogonalization of the set $\{\sum_i T_i, T_1,\ldots,T_{d^2-1}\}$ starting from the first entry. 
Then, 
\begin{equation*}
A_i := \frac{1}{\sqrt{d}} \sum_{j=0}^{d^2-1} \Bracket{S_j}{T_i }T_j \quad (i=0,1,\ldots,d^2-1) 
\end{equation*}
is a COB. }

Note that we can obtain an explicit formula for this construction as 
\begin{subequations}
\begin{eqnarray*}
A_0 &=& \frac{1}{d\sqrt{d}}\Bigl(T_0 - \sum_{j=1}^{d^2-1} f(j) T_j \Bigl), \\
A_i &=&  \frac{1}{d\sqrt{d}}\Bigl(T_0 - \sum_{j=1}^{i-1} f(j) T_j  + (d^2-i)f(i) T_i \Bigr) \\
&& (i=1,2\ldots,D-1),
\end{eqnarray*}
\end{subequations}
where $f(j) = \frac{d}{\sqrt{(d^2-j)(d^2-(j-1))}}$. 

\begin{figure}[h]
\begin{center}
\includegraphics[width=8cm]{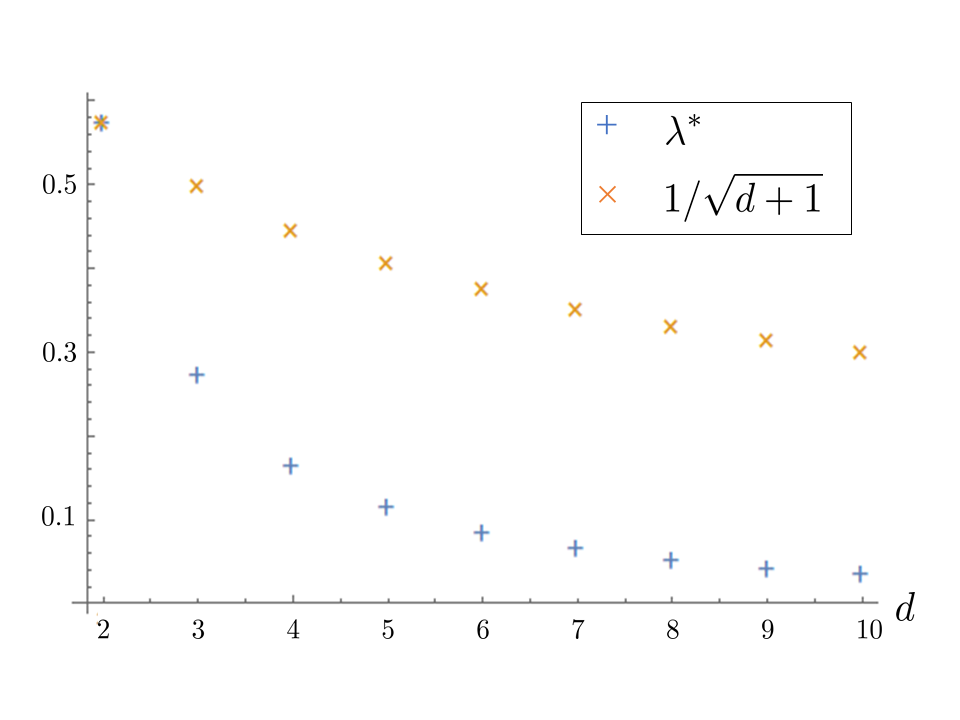}
\end{center}
\caption{Value $\lambda^\ast$ of COBs made by Construction 2 and the optimal value $\frac{1}{\sqrt{d+1}}$}
\label{NumFIG}
\end{figure}

The canonical construction \eqref{DefGk} for $d=2$ using the standard Pauli matrices gives a SIC POVM 
\begin{eqnarray*}
G_0 &=& 
\frac{1}{12}
  \left(
    \begin{array}{cc}
      -\sqrt{6}+3 & -1+\sqrt{2}i  \\
      -1-\sqrt{2}i & \sqrt{6}+3 
    \end{array}
  \right), \\
G_1 &=& 
\frac 1 4
  \left(
    \begin{array}{cc}
      1 & 1 \\
      1 & 1 
    \end{array}
  \right), \\
G_2 &=& 
\frac{1}{12}
  \left(
    \begin{array}{cc}
      3 & -1-2\sqrt{2}i  \\
      -1+2\sqrt{2}i & 3 
    \end{array}
  \right), \\
G_3 &=& 
\frac{1}{12}
  \left(
    \begin{array}{cc}
      \sqrt{6}+3 & -1+\sqrt{2}i  \\
      -1-\sqrt{2}i & -\sqrt{6}+3 
    \end{array}
  \right). 
\end{eqnarray*}
For a general $d \ge 3$, we can also compute the COB using the generalized Gell-Mann matrices: 
\begin{eqnarray}
T_{nm} &:=& 
\left\{
\begin{array}{l}
\frac{1}{\sqrt{2}}(\ketbra{n}{m} + \ketbra{m}{n}) \quad  (n<m), \\
\frac{i}{\sqrt{2}}(\ketbra{n}{m} - \ketbra{m}{n}) \quad (n>m),
\end{array}
\right.  \nonumber\\
T_{nn} &:=& \frac{1}{n\sqrt{n+1}}\Bigg( \sum_{k=1}^n \ketbra{k}{k} - n\ketbra{n+1}{n+1} \Bigg)  \nonumber\\
&& (n=1,2,\ldots ,d-1), \label{GM_mat}
\end{eqnarray} 
and $T_{dd} := \frac{1}{\sqrt{d}}\I$. 
We have numerically computed $\tau$ of the COB and plotted $\lambda^\ast$ in Fig. \ref{NumFIG}. 
Except for $d=2$, $\lambda^\ast$ is less than the maximum value $\frac{1}{\sqrt{d+1}}$, so the corresponding GSIC POVMs are not SIC POVMs. 

The third construction is based on the complete sets of mutually unbiased bases (MUBs) and mutually unbiased striations (MUSs) \cite{Iv81,WF88}. 
Let us first give a short review of those concepts. 

Two orthonormal bases (ONBs) $(\ket{\psi_i})_{i=1}^d$ and $(\ket{\phi_i})_{i=1}^d$ for $\HA$ are called mutually unbiased if $|\bracket{\psi_i}{\phi_j}|^2 = \frac 1 d$ for all $i,j$. 
The set of ONBs $(\ket{J,i})_{i=1}^d \ (J = 1,2,\ldots ,m)$ is called mutually unbiased if any pair of the bases is mutually unbiased: 
\begin{equation*}
|\bracket{J,i}{J',i'}|^2 = \delta_{JJ'}\delta_{ii'} + \frac{1}{d}(1-\delta_{JJ'}). 
\end{equation*} 
The maximum number of MUBs is known to be $d+1$ and the set of MUBs with $d+1$ elements is called complete. 
Similar to the problem of SIC POVMs, the existence of the complete set of MUBs for all $d$ is still open. 

Next let $M$ denotes a set with the cardinality $\# (M) = d^2$, which we label as $M = \{1,2,\ldots,d^2\}$. 
A subset of $M$ is called a line. 
A set of $d$ lines $( L_i)_{i=1}^d (L_i\subset M)$ 
is called striations of $M$ if $\# (L_i\cap L_j) =d\delta_{ij}$ holds. 
Since $\# (M) = d^2$, the set of striations $( L_i)_{i=1}^d$ forms a partition of $M$. 
Two striations $( L_i)_{i=1}^d$ and $( K_i)_{i=1}^d$ are called mutually unbiased striations (MUSs) if $\# (L_i \cap K_{j}) = 1$ for all $i,j$ \cite{Wo06}. 
The set of striations $( L^{(J)}_i)_{i=1}^d$ ($J=1,2,\ldots ,m'$) is called mutually unbiased (or the orthogonal Latin squares) if any pair of the striations is mutually unbiased:  
\begin{equation*}
\# \Big(L_i^{(J)}\cap L_{i'}^{(J')}\Big) =d\delta_{JJ'}\delta_{ii'}+(1-\delta_{JJ'}). 
\end{equation*}
The maximum number of MUSs is known to be $d+1$ and the set of MUSs with $d+1$ elements is called complete. 

With these similarities between MUBs and MUSs in mind, Wootters showed the followings. 
Let $(A_i)_{i=1}^{d^2}$ be a COB, $(L_i^{(J)})_{i=1}^d$ a complete set of MUSs of $M$, and $( \ket{J,i})_{i=1}^d$ a set of ONBs for $\HA$ ($J=1,2,\ldots ,d+1$). 
If the equation 
\begin{equation*}
\ketbra{J,i}{J,i} = \sum_{k\in L_i^{(J)}}A_k \quad \forall J,i 
\end{equation*}
holds, then the set of bases $(\ket{J,i})_{i=1}^d$ is a complete set of MUBs. 

The following result shows the converse is also true and gives a construction of a COB by using complete sets of MUBs and MUSs. 

\noindent {\bf Construction 3.} {\it 
Let $( \ket{J,i})_{i=1}^d$ and $(L_i^{(J)})_{i=1}^d$ ($J=1,2,\ldots,d+1$) be complete sets of MUBs and MUSs. 
Define the function $s : \{1,2,\ldots,d^2\} \times \{1,2,\ldots,d+1\} \to \{1,2,\ldots,d\}$ by $s(k,J):=i$ such that $k\in L_i^{(J)}$. 
Notice that such a function uniquely exists since $(L^{(J)}_i)_{i=1}^{d}$ for each $J$ forms a partition of $\{1,\ldots,d^2\}$. 
Then 
\begin{eqnarray} 
A_k &=&\frac{1}{d}\Bigg(\sum_{J=1}^{d+1}\ketbra{J,s(k,J)}{J,s(k,J)} - \I\Bigg) \label{Ak} \\
&&(k=1,2,\ldots ,d^2) \nonumber
\end{eqnarray}
is a COB, as is shown below. }

Following \cite{HHH205}, we introduce a vector $\ket{\Phi_{J,i}} := \overline{\ket{J,i}}\otimes \ket{J,i}$ on $\HA \otimes \HA$ 
where $\overline{\ket{\psi}} := \sum_i \overline{\bracket{i}{\psi}} \ket{i}$ denotes the complex conjugate vector with respect to a (fixed) ONB $(\ket{i})_{i=1}^d$. Let $\ket{\Psi} := \frac{1}{\sqrt{d}} \sum_{i=1}^d \ket{i}\otimes\ket{i}$ be a maximally entangled state. 
Then it is easy to see that $\ket{\Psi} = \frac{1}{\sqrt{d}} \sum_{i=1}^d \overline{\ket{J,i}}\otimes\ket{J,i}$ for any $J$. 
Let 
\begin{equation}\label{kvec}
\ket{\hat k}:=\frac{1}{\sqrt d} \sum_{J=1}^{d+1}\ket{\Phi_{J,s(k,J)}}-\ket{\Psi} \quad (k=1,2,\ldots ,d^2). 
\end{equation}
Then, one can show that $\ket{\hat k}$ is a unit vector and $\bracket{\Phi_{J,i}}{\hat k} = \frac{1}{\sqrt d}\delta_{i,s(k,J)}$. 
Moreover, one can show $\sum_i \ketbra{\hat k}{\hat k} = \I$ by the completeness conditions for MUBs and MUSs; 
hence $\{\ket{\hat k}\}_{k=1}^{d^2}$ forms an ONB for $\HA\otimes \HA$ (see \cite{HHH205} for details). 

Now consider an isomorphism $A\in\LA(\HA) \mapsto \IA(A):=\ket{(\I\otimes A)\Psi} \in \HA\otimes\HA$ between an operator and a vector. 
As $(\ket{i})_{i=1}^{d}$ forms a basis for $\HA$, it is easy to see that $\IA$ is a linear bijection between $\LA(\HA)$ and $\HA\otimes \HA$, and $\Bracket{A}{B} = d \bracket{\IA(A)}{\IA(B)}$ for any $A,B\in\LA (\HA)$. 
We define $A_k\in\LA(\HA)$ ($k=1,2,\ldots ,d^2$) by 
\begin{equation}\label{DefAk}
\IA(A_k)= \frac{1}{d}\ket{\hat k} 
\end{equation}
Then the normalization condition holds:
\begin{equation*}
\Bracket{A_k}{A_{k'}} 
= d \bracket{\IA(A_k)}{\IA(A_{k'})} 
= \frac{\bracket{\hat k}{\hat k'} }{d}
= \frac{\delta_{kk'}}{d}.
\end{equation*}
Noting that $\ket{\Phi_{J,i}} = \sqrt{d} \IA(\ketbra{J,i}{J,i})$, 
we have $\Bracket{A_k}{\ketbra{J,i}{J,i}} = d \bracket{\IA(A_k)}{\IA(\ketbra{J,i}{J,i})} = \frac{1}{\sqrt{d}} \bracket{k}{\Phi_{J,i}} = \frac{1}{d} \delta_{i,s(k,J)}$.    
So, we observe $\ketbra{J,i}{J,i} = d\sum_{k=1}^{d^2}\Bracket{A_k}{\ketbra{J,i}{J,i}}A_k =\sum_{k\in L_i^{(J)}}A_k$. 
Then, 
\begin{equation*}
\I = \sum_i \ketbra{J,i}{J,i}= \sum_i \sum_{k \in L^{(J)}_i} A_k = \sum_{k=1}^{d^2} A_k. 
\end{equation*}
Hence, the set $(A_k)_k$ is a COB. 

Finally, the explicit form of $A_k$ is shown as follows.  
We have $\IA (\ketbra{\phi}{\phi})=\frac{1}{\sqrt d}\overline{\ket{\phi}}\otimes\ket{\phi}$ for any $\ket{\phi}$ and $\IA (\I)=\ket{\Psi}$. 
By using these properties, as well as \eqref{kvec}, and \eqref{DefAk}, one arrives at the expression \eqref{Ak}.  

Let us construct a SIC POVM for $d=2$ using Construction 3. 
We employ the sets 
$
L_1^{(1)}:=\{1,2\}, L_2^{(1)}:=\{3,4\}, 
L_1^{(2)}:=\{1,3\}, L_2^{(2)}:=\{2,4\}, 
L_1^{(3)}:=\{1,4\}
$
, and 
$
L_2^{(3)}:=\{2,3\} 
$ 
as a complete set of MUSs and the set of bases 
$
( |1,1\rangle :=\frac{1}{\sqrt{2}} (1,1)^T,|1,2\rangle :=\frac{1}{\sqrt{2}} (1,-1)^T), 
( |2,1\rangle :=\frac{1}{\sqrt{2}} (1,i)^T,|2,2\rangle :=\frac{1}{\sqrt{2}} (1,-i)^T)
$
, and 
$
( |3,1\rangle :=(1,0)^T,|3,2\rangle :=(0,1)^T)
$ 
as a complete set of MUBs. 
According to the direct computation using \eqref{Ak}, a canonical construction \eqref{DefGk} gives the following SIC POVM: 
\begin{eqnarray*}
G_1 &=& \frac{1}{4\sqrt{3}}
\begin{pmatrix}
1+\sqrt 3 & 1-i\\
1+i & -1+\sqrt 3
\end{pmatrix}
,
\\
G_2 &=& \frac{1}{4\sqrt{3}}
\begin{pmatrix}
-1+\sqrt 3 & 1+i\\
1-i & 1+\sqrt 3
\end{pmatrix}
,
\\
G_3 &=& \frac{1}{4\sqrt{3}}
\begin{pmatrix}
-1+\sqrt 3 & -1-i\\
-1+i & 1+\sqrt 3
\end{pmatrix}
,  
\\
G_4 &=& \frac{1}{4\sqrt{3}}
\begin{pmatrix}
1+\sqrt 3 & -1+i\\
-1-i & -1+\sqrt 3
\end{pmatrix}
.
\end{eqnarray*}

\section{Conclusions and discussions}\label{sum}
In this paper we gave the construction of GSIC POVM s by means of COBs and investigated the condition to give a SIC POVM by the spectrum property of a COB (Theorem \ref{const:genSIC}). 
In particular, for $d=2$, any canonically constructed GSIC POVM is a SIC POVM (Proposition \ref{prop:t1}), while for $d \ge 3$, conditions for the power of traces of a COB were given to yield SIC POVMs (Proposition \ref{prop:t2}). 
A characteristic value $\lambda$ of a COB gives the bound of the scaled MSE for the linear quantum state tomography by using IC POVMs. 
We then provided three different constructions of COBs, one of which shows a relation to MUBs.  
The constructions serve not only for GSIC POVMs, but also for NQPRs. 

Finally, we offer another idea of construction of COBs, and hence of GSIC POVMs, based on Zauner's conjecture for a SIC POVM \cite{ref:Zau,RBSC04}. 
Let $(D_{jk})_{j,k=0}^{d-1}$ be the tuple of unitary operators defined by 
\begin{equation*}
D_{jk} = \omega^{\frac{jk}{2}} \sum_{m=0}^{d-1} \omega^{jm} \ketbra{k \oplus m}{m},
\end{equation*}
where $\omega = \exp(\frac{2\pi i}{d})$, $(\ket{k})_k$ is an ONB for $\HA$, and $\oplus$ denotes the addition modulo $d$. 
Then, it is believed that there is a normalized fiducial vector $\ket{\phi} \in \HA$ with which $(\ketbra{\psi_{jk}}{\psi_{jk}})_{j,k=0}^{d-1}$ where $\ket{\psi_{jk}} = \frac{1}{\sqrt{d}} D_{jk} \ket{\phi}$ is a SIC POVM. 
Note that $(D_{jk})_{j,k}$ is a faithful projective unitary representation of a group $G = Z_d \times Z_d$. 
More generally, for a group $G$ with the identity $e$ and the order $\# (G) = d^2$, let $(U_g)_{g \in G}$ be a faithful projective unitary representation:  
\begin{equation}\label{eq:PrUni}
U_g U_{g'} = c(g,g') U_{g g'} \quad (g,g' \in G)
\end{equation}
with $|c(g,g')| = 1$ which is orthogonal 
\begin{equation}\label{eq:orthU}
\Bracket{U_g}{U_{g'}} = d \delta_{g g'}. 
\end{equation}
These bases are sometimes called nice error bases \cite{ref:NEB,ref:NEB2}. 
Note that the faithfullness is required to guarantee $\# (U_g) = d^2$. 
By the properties \eqref{eq:PrUni} and $|c(g,g')| = 1$, it is easy to see that if a fiducial vector satisfies 
\begin{equation}\label{eq:fidCond}
|\bracket{\phi}{U_g \phi}|^2 = \frac{1}{d+1} \quad \forall g \neq e, 
\end{equation}
$(\frac{1}{d} \ketbra{U_g \phi}{U_g \phi})_{g \in G}$ forms a SIC POVM. 
Note that \cite{ref:Wer} the orthogonality condition \eqref{eq:orthU} is equivalent to the relation
\begin{equation}\label{eq:UCUd}
\sum_g U_g C U^\dagger_g = d (\tr C) \I \quad \forall C\in\LA (\HA); 
\end{equation}
hence the completeness of the POVM follows automatically. 

Let $(U_g)_{g \in G}$ be a faithful projective unitary representation of a group $G$ with $\# (G) =d^2$. 
Let $A$ be an Hermitian operator with $\tr A^2 = \frac{1}{d}$. Moreover, let $A$ satisfies the condition  
\begin{equation*}
\tr A U_g^\dagger A U_g = 0 \quad \forall g \neq e. 
\end{equation*} 
Then it is easy to see that $(A_g := U_g A U_g^\dagger)_{g \in G}$ is a COB: 
The orthogonality and the completeness conditions follow from \eqref{eq:fidCond} and \eqref{eq:UCUd}, respectively. 
Note also that $\tr A_g = \frac{1}{d}$ follows automatically. Such an operator $A$ might be called a fiducial operator. 
Hence, a construction for both SIC and GSIC POVMs reduces to the problem of finding a fiducial operator. 
We think the problem is interesting even for GSIC POVMs.  

\appendix

\begin{acknowledgements}
G. K. was supported in part by JSPS KAKENHI Grant No. 17K18107. 
\end{acknowledgements}

\section{Proofs of some propositions}\label{ProofLem}
In this appendix we give proofs of some propositions and lemmas. 

\subsection{Spanning property of IC-POVM}\label{spICPOVM}

First, the following is a well-known fact for the IC POVM (see, e.g., \cite{DPS04}), but here we provide its simple proof.  
\begin{proposition}\label{thm:sp}
A POVM $F = (F_i)_{i=1}^n $ is informationally complete if and only if $F$ spans $\LA(\HA)$. 
\end{proposition}
{\it Proof.} \ 
Note first that $F$ is informationally complete if and only if for any $C,D \in \LA(\HA)$ and any $i$, $\  \tr F_i C = \tr F_i D \Rightarrow C = D$ by noting that any linear operator can be expressed as a linear combination of density operators.  

Let $F = (F_i)_{i=1}^n $ be an informationally complete POVM. 
Assume the contrary, that $F$ does not span $\LA(\HA)$. 
Then $({\rm span} F)^{\perp} \neq \{0\}$. 
Namely, there is non zero $X \in \LA(\HA)$ such that for any $i$, $\Bracket{F_i}{X} = \tr F_i X = 0 = \tr (F_i 0)$. 
However, the IC POVM then implies $X = 0$, which is a contradiction. 
The converse is trivial. 
\hfill $\blacksquare$ 

\bigskip 

\subsection{Proof of Lemma \ref{lem:xd}}\label{pr:xd}

{\it Proof of Lemma \ref{lem:xd}}. Conditions (i) and (ii) imply that $x_1 > 0$ and $x_d < 0$: 
To see this, assume contrary that $x_d \ge 0$, so that all $x_i \ge 0$. 
Then, $(\sum_{i}x_i)^2 - (\sum_i x^2_i) = \sum_{i \neq j} x_i x_j \ge 0$. 
On the other hand, by (i) and (ii), $(\sum_{i}x_i)^2 - (\sum_i x^2_i) = (\frac{1}{d})^2 - \frac{1}{d} = \frac{1-d}{d^2} < 0$. 
Thus we have a contradiction. 
A similar argument (by flipping the sign) shows $x_1 > 0$. 

Let $a_i := \frac{\sqrt{2}}{\sqrt{d(d^2-1)}} (d^2 x_i - 1)$. 
It follows from (i) and (ii) that 
\begin{equation*}
(i)' \ \sum_i a_i = 0, \quad (ii)' \ \sum_i a_i^2 = 2. 
\end{equation*}
Similar to the above argument, $a_d < 0$ and one sees $|a_d| =  \frac{\sqrt{2}}{\sqrt{d(d^2-1)}} (d^2 |x_d| + 1)$ (note that $x_d < 0$ implies $|1 - d^2 x_d| = d^2 |x_d| + 1$). 
Proposition 1-[I] in \cite{ref:KK} shows 
\begin{equation}\label{ad_bound} 
|a_d| \ge \sqrt{\frac{2}{d(d-1)}}. 
\end{equation}
Therefore, we have 
\begin{equation*}
\frac{\sqrt{2}}{\sqrt{d(d^2-1)}} (d^2 |x_d| + 1) \ge \sqrt{\frac{2}{d(d-1)}}, 
\end{equation*}
from which we obtain \eqref{eq:ineq}. 
By Proposition 1-[III] in \cite{ref:KK}, \eqref{ad_bound} is saturated, which implies \eqref{eq:ineq} is saturated, if and only if 
\begin{eqnarray*}
a_1  &=& \sqrt{\frac{2(d-1)}{d}}, \\
a_2,a_3,\ldots , a_d &=& -\sqrt{\frac{2}{d(d-1)}}, 
\end{eqnarray*}
which imply that \eqref{x1} and \eqref{x2_d} hold. 
\hfill $\blacksquare$

\subsection{Proof of \eqref{Tau_Neg}}\label{pr:TauNeg}
{\it Proof of \eqref{Tau_Neg}}. 
We denote by $m_i$ the minimum eigenvalue of $A_i$, which is strictly negative as is shown in the main text. 
Clearly, $-|m_k|= m_k \le \tr \rho A_k $ for any $\rho \in \SA(\rho)$. 
Hence, we have $\tau = \max_i \{|m_i|\} \ge - \tr \rho A_k$ 
and thus $\tau = \max_i \{|m_i|\} \ge - \min_k \tr \rho A_k$. 
Since the positivity of $\tau$ trivially holds, this shows that 
\begin{equation*}
\frac{1}{d} N(\{A_i\}) \le \tau.  
\end{equation*}
To prove the converse inequality, let $\rho_k = \ketbra{\phi_k}{\phi_k}\in \SA(\HA)$ where $\ket{\phi_k}$ is the unit eigenvector of $A_k$ corresponding to the minimum eigenvalue $m_k$. 
We have, for any $k$, 
\begin{equation*}
\min_i \tr A_i \rho_k = \min_i \bracket{\phi_k}{A_i \phi_k} \le \bracket{\phi_k}{A_k \phi_k} = m_k.  
\end{equation*}
Therefore, $\frac{1}{d} N(\{A_i\}) \ge \max\{0, - \min_i \tr A_i \rho_k\} \ge - \min_i \tr A_i \rho_k \ge -m_k = |m_k|$. Since this holds for any $k$, we have 
\begin{equation*}
\frac{1}{d} N(\{A_i\}) \ge  \tau.  
\end{equation*}

\section{Orthogonal basis with a fixed sum}\label{app:obwfs}
Let $\KA$ be a $D$-dimensional real inner product space. 
In this appendix, we provide two constructions of an orthogonal basis $\{\ket{\phi_i}\}_{i=0}^{D-1}$ with constant norms, i.e., $\bracket{\phi_i}{\phi_j} = c \delta_{ij}$ ($c > 0$), as well as the fixed sum $\sum_{i=0}^{D-1} \ket{\phi_i} = \ket{\iota}$. 
Note that automatically $c = \frac{\|\iota\|^2}{D}$ since $\|\iota\|^2 = \bracket{\sum_{i} \phi_i}{\sum_j \phi_j} = \sum_{i,j} c \delta_{ij} = c D$. 
For our purpose of constructing a COB, just apply $\ket{\iota} = \I$ where $\KA$ is the real Hilbert space of Hermitian operators, noting that $\|\I\| = \sqrt{d}$ and $D = d^2$. 

\bigskip 

\noindent {\bf Construction 1}. {\it 
With a given $\ket{\iota}$, prepare an orthonormal basis $\{\ket{t_i}\}_{i=0}^{D-1}$ where $\ket{t_0} = \frac{\ket{\iota}}{\|\iota\|}$. 
Prepare also an orthogonal $D \times D$ real matrix $O = [O_{ij}]_{i,j=0}^{D-1}$ (i.e., $O O^T = O^T O = I$) such that 
\begin{equation}
O_{0i} = \frac{1}{\sqrt{D}} \quad \forall i = 0,1,\ldots,D-1, \label{eq:subO2}
\end{equation}
where $I$ is the $D \times D$ identity matrix. 
Then, 
\begin{equation}\label{eq:const1}
\ket{\phi_i} := \frac{\|\iota\|}{\sqrt{D}} \sum_{j=0}^{D-1} O_{ji} \ket{t_j} \quad (i=0,1,\ldots,D-1)
\end{equation}
gives the desired basis.} 

Indeed, since $O$ is an orthogonal matrix, one has $\ket{t_i} = \frac{\sqrt{D}}{\|\iota\|} \sum_j O_{ij} \ket{\phi_j}$ so that the condition \eqref{eq:subO2} implies 
\begin{equation*}
\ket{\iota} = \|\iota\| \ket{t_0} = \sum_j \sqrt{D}  O_{0j} \ket{\phi_j} = \sum_j \ket{\phi_j}.
\end{equation*}
The orthogonality of $\ket{\phi_i}$ is also satisfied by the orthogonality of $O$. 

Note here that both $\{\ket{t_i}\}_i$ and $O$ can be easily constructed by using the Gram-Schmidt orthogonalization starting from $\ket{\iota}$ and $(1,1,\ldots,1)^T \in \R^D$, respectively. 

Note also that, conversely, any orthogonal basis $\{\ket{\phi_i}\}_{i=0}^{D-1}$ with a fixed sum $\ket{\iota}$ can be constructed in this way [with two alternatives (i) and (ii) below].  

(i) Given an arbitrary orthonormal basis $\{\ket{t_i}\}_i$ with $\ket{t_0} = \frac{\ket{\iota}}{\|\iota\|}$, there exists an orthogonal matrix $O$ satisfying \eqref{eq:subO2} such that any orthogonal basis $\{\ket{\phi_i}\}_{i=0}^{D-1}$ with a fixed sum $\ket{\iota}$ is constructed by \eqref{eq:const1}.
\footnote{
As $\{\ket{t_i}\}_i$ and $\{\ket{\phi_j}\}_j$ are both orthogonal bases, there exists an orthogonal matrix $O=[O_{ij}]$ which connects them: $\ket{t_i} = \frac{\sqrt{D}}{\|\iota\|}\sum_{j}O_{ij}\ket{\phi_j}$. 
Since $\frac{1}{\|\iota\|}\sum_{j}\ket{\phi_j}=\ket{t_0}=\frac{\sqrt{D}}{\|\iota\|}\sum_{j}O_{0j}\ket{\phi_j}$, 
one has $O_{0j}=\frac{1}{\sqrt{D}}$ for all $j$. 
} 

(ii) Given an arbitrary orthogonal matrix $O$ satisfying \eqref{eq:subO2}, there exists $\{\ket{t_i}\}_i$ with $\ket{t_0} = \frac{\ket{\iota}}{\|\iota\|}$ such that any orthogonal basis $\{\ket{\phi_i}\}_{i=0}^{D-1}$ with a fixed sum $\ket{\iota}$ is constructed by \eqref{eq:const1}
\footnote{Let $O=[O_{ij}]$ be an orthogonal matrix satisfying \eqref{eq:subO2}. 
Then it is straightforward to see that $\ket{t_i}=\frac{\sqrt{D}}{\|\iota\|}\sum_jO_{ij}\ket{\phi_j}$ is the desired basis. }. 

The next construction is not general but uses only one orthonormal basis and is more concrete.  

\noindent {\bf Construction 2}. 
Prepare an orthonormal basis $\{\ket{t_i}\}_{i=0}^{D-1}$ where $\ket{t_0} = \frac{\ket{\iota}}{\|\iota\|}$. 
Construct an orthonormal basis $\{\ket{s_i}\}_{i=0}^{D-1}$ by the Gram-Schmidt orthogonalization of the set $S = \{\ket{s}, \ket{t_1},\ldots,\ket{t_{D-1}}\}$ where $\ket{s} := \sum_j \ket{t_j}$ starting from $\ket{s}$. 

Then, using the unitary operator $U = \sum_{j} \ketbra{t_j}{s_j}$, it is easy to see that 
\begin{equation}\label{eq:const2}
\ket{\phi_i} := \frac{\|\iota\|}{\sqrt{D}} U \ket{t_i}
\end{equation}
gives a desired orthogonal basis. In particular, $\sum_i \ket{\phi_i} = \frac{\|\iota\|}{\sqrt{D}} U \ket{s} = \sum_i \ket{t_i}$ since $\ket{s_0} = \frac{\ket{s}}{\|s\|} = \frac{1}{\sqrt{D}}\sum_i \ket{t_i}$. 

Note that the linear independence of the set $S$ is easily shown. 
The choice of the latter $D-1$ vectors in $S$ can be arbitrary from $\{\ket{t_i}\}_{i=0}^{D-1}$. 
However, by the symmetric argument, one can show that the obtained orthonormal basis $\{\ket{s_i}\}_{i=0}^{D-1}$ is independent of the choice. 

One can continue this construction more concretely as follows. 
First, the direct computation of the Gram-Schmidt orthogonalization gives $\ket{s_0} = \frac{1}{\sqrt{D}}\sum_j \ket{t_j}$ and 
\begin{equation*}
\ket{s_i} = \frac{(D-i)\ket{t_i} - \sum_{k \neq 1,2,\ldots,i}\ket{t_k}}{\sqrt{(D-i)(D-(i-1))}} \quad  (1\leq i\leq D-1). 
\end{equation*}
Plugging this into \eqref{eq:const2}, one arrives at the COB given by 
\begin{eqnarray*}
\ket{\phi_0} &=& \frac{\|\iota\|}{D}\Bigl(\ket{t_0} - \sum_{j=1}^{D-1} f(j) \ket{t_j} \Bigl) \\
 \ket{\phi_i} &=&  \frac{\|\iota\|}{D}\Bigl(\ket{t_0} - \sum_{j=1}^{i-1} f(j) \ket{t_j}  + (D-i)f(i) \ket{t_i} \Bigr) \\
&& (i=1,2,\ldots,D-1) 
\end{eqnarray*}
where $f(j) = \frac{\sqrt{D}}{\sqrt{(D-j)(D-(j-1))}}$. 

Finally, here are some examples of COBs. 
In $d=2$, 
Construction 1 using 
$T_0=\frac{1}{\sqrt{2}}\I,T_1=\frac{1}{\sqrt{2}}\sigma_x,T_2=\frac{1}{\sqrt{2}}\sigma_y,T_3=\frac{1}{\sqrt{2}}\sigma_z$, and 
\begin{eqnarray*}
O =
\begin{pmatrix}
 \frac{1}{2} & \frac{1}{2} & \frac{1}{2} &\frac{1}{2} \\
 \frac{1}{2} & -\frac{1}{2} & \frac{1}{2} &-\frac{1}{2} \\
 \frac{1}{2} & \frac{1}{2} & -\frac{1}{2} &-\frac{1}{2} \\
 \frac{1}{2} & -\frac{1}{2} & -\frac{1}{2} &\frac{1}{2} \\
\end{pmatrix} 
\end{eqnarray*} 
gives the following COB: 
\begin{eqnarray}
&&
A_1 =
\begin{pmatrix}
  \frac{1}{2} & \frac{1-i}{4} \\
 \frac{1+i}{4} & 0 \\
\end{pmatrix}, 
A_2 =
\begin{pmatrix}
 0 & \frac{-1-i}{4} \\
 \frac{-1+i}{4} &  \frac{1}{2} \\
\end{pmatrix}, \nonumber
\\ 
&&
A_3 =
\begin{pmatrix}
 0 & \frac{1+i}{4} \\
 \frac{1-i}{4} &  \frac{1}{2} \\
\end{pmatrix},
A_4 =
\begin{pmatrix}
  \frac{1}{2} & \frac{-1+i}{4} \\
 \frac{-1-i}{4} & 0 \\
\end{pmatrix}.\label{ex_const1}
\end{eqnarray}

Construction 2 using the above $(T_i)_{i=0}^3$ gives the following COB: 
\begin{eqnarray*}
A_1 &=&
\begin{pmatrix}
 \frac{1}{4}-\frac{1}{2 \sqrt{2}} & -\frac{1}{4 \sqrt{3}}+\frac{i}{2 \sqrt{6}} \\
 -\frac{1}{4 \sqrt{3}}-\frac{i}{2 \sqrt{6}} & \frac{1}{4}+\frac{1}{2 \sqrt{2}} \\
\end{pmatrix}, 
\\
A_2 &=&
\begin{pmatrix}
 \frac{1}{4} & \frac{\sqrt{3}}{4} \\
 \frac{\sqrt{3}}{4} & \frac{1}{4} \\
\end{pmatrix},
\\
A_3 &=&
\begin{pmatrix}
 \frac{1}{4} & -\frac{1}{4 \sqrt{3}}-\frac{i}{\sqrt{6}} \\
 -\frac{1}{4 \sqrt{3}}+\frac{i}{\sqrt{6}} & \frac{1}{4} \\
\end{pmatrix},
\\
A_4 &=&
\begin{pmatrix}
 \frac{1}{4}+\frac{1}{2 \sqrt{2}} & -\frac{1}{4 \sqrt{3}}+\frac{i}{2 \sqrt{6}} \\
 -\frac{1}{4 \sqrt{3}}-\frac{i}{2 \sqrt{6}} & \frac{1}{4}-\frac{1}{2 \sqrt{2}} \\
\end{pmatrix}.
\end{eqnarray*} 

Construction 3 using 
$
L_1^{(1)}=\{1,2\}, L_2^{(1)}=\{3,4\}, 
L_1^{(2)}=\{1,3\}, L_2^{(2)}=\{2,4\}, 
L_1^{(3)}=\{1,4\}
$
, and 
$
L_2^{(3)}=\{2,3\} 
$ 
and a complete set of MUBs which consists of the normalized eigenvectors of the Pauli matrices gives the same COB as in \eqref{ex_const1}. 

In $d=2$, as mentioned in the proof of Proposition \ref{prop:t1}, 
all eigenvalues of $A_i$ are the same $\frac{1\pm \sqrt{3}}{4}$. 
Therefore, we observe $\lambda^{\ast} = \frac{1}{\sqrt{3}}$, which can saturate the upper bound. 

In three or more dimensions, the matrix forms of COBs are more complex. 
For example, Construction 2 for $d=3$ using the generalized Gell-Mann matrices \eqref{GM_mat} gives the following COB: 
\clearpage
\begin{widetext}
\begin{eqnarray*}
A_1 &=& 
\begin{pmatrix}
 -\frac{2}{9} & \frac{-7+3 i \sqrt{7}}{84 \sqrt{3}} & \frac{i}{6 \sqrt{5}}-\frac{1}{6 \sqrt{7}} \\
 \frac{-7-3 i \sqrt{7}}{84 \sqrt{3}} & \frac{1}{9} & -\frac{-5 i+\sqrt{15}}{30 \sqrt{2}} \\
 -\frac{i}{6 \sqrt{5}}-\frac{1}{6 \sqrt{7}} & -\frac{5 i+\sqrt{15}}{30 \sqrt{2}} & \frac{4}{9} \\
\end{pmatrix}, \quad 
A_2 = 
\begin{pmatrix}
 \frac{1}{9} & \frac{2}{3 \sqrt{3}} & 0 \\
 \frac{2}{3 \sqrt{3}} & \frac{1}{9} & 0 \\
 0 & 0 & \frac{1}{9} \\
\end{pmatrix}, \quad 
A_3 = 
\begin{pmatrix}
 \frac{1}{9} & \frac{-1-3 i \sqrt{7}}{12 \sqrt{3}} & 0 \\
 \frac{-1+3 i \sqrt{7}}{12 \sqrt{3}} & \frac{1}{9} & 0 \\
 0 & 0 & \frac{1}{9} \\
\end{pmatrix}, \\
A_4 &=& 
\begin{pmatrix}
 \frac{1}{9} & \frac{-7+3 i \sqrt{7}}{84 \sqrt{3}} & \frac{1}{\sqrt{7}} \\
 \frac{-7-3 i \sqrt{7}}{84 \sqrt{3}} & \frac{1}{9} & 0 \\
 \frac{1}{\sqrt{7}} & 0 & \frac{1}{9} \\
\end{pmatrix},\quad 
A_5 = 
\begin{pmatrix}
 \frac{1}{9} & \frac{-7+3 i \sqrt{7}}{84 \sqrt{3}} & -\frac{i \sqrt{5}}{6}-\frac{1}{6 \sqrt{7}} \\
 \frac{-7-3 i \sqrt{7}}{84 \sqrt{3}} & \frac{1}{9} & 0 \\
 \frac{i \sqrt{5}}{6}-\frac{1}{6 \sqrt{7}} & 0 & \frac{1}{9} \\
\end{pmatrix}, \\
A_6 &=& 
\begin{pmatrix}
 \frac{1}{9} & \frac{-7+3 i \sqrt{7}}{84 \sqrt{3}} & \frac{i}{6 \sqrt{5}}-\frac{1}{6 \sqrt{7}} \\
 \frac{-7-3 i \sqrt{7}}{84 \sqrt{3}} & \frac{1}{9} & \sqrt{\frac{2}{15}} \\
 -\frac{i}{6 \sqrt{5}}-\frac{1}{6 \sqrt{7}} & \sqrt{\frac{2}{15}} & \frac{1}{9} \\
\end{pmatrix}, \quad 
A_7 =
\begin{pmatrix}
 \frac{1}{9} & \frac{-7+3 i \sqrt{7}}{84 \sqrt{3}} & \frac{i}{6 \sqrt{5}}-\frac{1}{6 \sqrt{7}} \\
 \frac{-7-3 i \sqrt{7}}{84 \sqrt{3}} & \frac{1}{9} & -\frac{15 i+\sqrt{15}}{30 \sqrt{2}} \\
 -\frac{i}{6 \sqrt{5}}-\frac{1}{6 \sqrt{7}} & -\frac{-15 i+\sqrt{15}}{30 \sqrt{2}} & \frac{1}{9} \\
\end{pmatrix}, \\ 
A_8 &=& 
\begin{pmatrix}
 \frac{4}{9} & \frac{-7+3 i \sqrt{7}}{84 \sqrt{3}} & \frac{i}{6 \sqrt{5}}-\frac{1}{6 \sqrt{7}} \\
 \frac{-7-3 i \sqrt{7}}{84 \sqrt{3}} & -\frac{2}{9} & -\frac{-5 i+\sqrt{15}}{30 \sqrt{2}} \\
 -\frac{i}{6 \sqrt{5}}-\frac{1}{6 \sqrt{7}} & -\frac{5 i+\sqrt{15}}{30 \sqrt{2}} & \frac{1}{9} \\
\end{pmatrix}, \quad
A_9 = 
\begin{pmatrix}
 \frac{1}{9} & \frac{-7+3 i \sqrt{7}}{84 \sqrt{3}} & \frac{i}{6 \sqrt{5}}-\frac{1}{6 \sqrt{7}} \\
 \frac{-7-3 i \sqrt{7}}{84 \sqrt{3}} & \frac{4}{9} & -\frac{-5 i+\sqrt{15}}{30 \sqrt{2}} \\
 -\frac{i}{6 \sqrt{5}}-\frac{1}{6 \sqrt{7}} & -\frac{5 i+\sqrt{15}}{30 \sqrt{2}} & -\frac{2}{9} \\
\end{pmatrix}.
\end{eqnarray*}
\end{widetext}
We numerically observed $\tau = 0.291347$ and $\lambda^\ast = 0.276081$ which cannot saturate the upper bound $0.5$.

\end{document}